\documentclass[aps,prl,reprint,a4paper,superscriptaddress]{revtex4-1}

\usepackage[pdftex]{hyperref,color,graphicx}
\hypersetup{colorlinks=true,linkcolor=blue,citecolor=blue,filecolor=blue,urlcolor=blue}
\usepackage{amsfonts,amssymb,amsmath,siunitx}
\usepackage[english]{babel}

\newif\ifusebibfile
\usebibfiletrue   
\usebibfilefalse

\newcommand{\figref}[2]{\hyperref[#1]{\ref{#1}(#2)}}
\newcommand{\ket}[1]{\mbox{$\vert #1 \rangle$}}

\begin{document}
\textheight=24.4cm
\selectlanguage{english}

\title{Note: In-situ measurement of vacuum window birefringence by atomic spectroscopy}

\author{Andreas Steffen} 	
\author{Wolfgang Alt}
\author{Maximilian Genske}
\author{Dieter Meschede}
\author{Carsten Robens}
\author{Andrea Alberti}
\email{alberti@iap.uni-bonn.de}
\affiliation{Institut f\"ur Angewandte Physik, Universit\"at Bonn,
Wegelerstr.~8, D-53115 Bonn, Germany}
\date{\today}

\begin{abstract}
We present an in-situ method to measure the birefringence of a single vacuum window by means of microwave spectroscopy on an ensemble of cold atoms. Stress-induced birefringence can cause an ellipticity in the polarization of an initially linearly-polarized laser beam. The amount of ellipticity can be reconstructed by measuring the differential vector light shift of an atomic hyperfine transition. Measuring the ellipticity as a function of the linear polarization angle allows us to infer the amount of birefringence $\Delta n$ at the level of $\num{e-8}$ and identify the orientation of the optical axes. The key benefit of this method is the ability to separately characterize each vacuum window, allowing the birefringence to be precisely compensated in existing vacuum apparatuses.
\end{abstract}

\maketitle

Many experiments in quantum optics rely on an accurate control of the polarization of the laser beams \mbox{\cite{Monroe:1996,Mandel:2003,Lee:2007,choi2007elimination,2011Chicireanu,Steffen:2012}}. Optical access of laser beams to ultrahigh vacuum apparatus is offered by vacuum windows, which are, in general, affected by stress-induced birefringence occurring after mounting and bake-out. 
While the typical values of the induced birefringence $\Delta n$ are in the order of $\num{e-6}$, values significantly below this magnitude require special attention in mounting the vacuum viewports to avoid deformations \cite{solmeyerVacuum}.
It is, thus, important for precision applications to be able to charac\-terize the amount of birefringence of each individual window. Knowing the amount of birefringence and the orientation of the principal axes makes it possible to avoid polarization distortions either by aligning the incoming linear polarization onto one of the optical axis, or by fully compensating the birefringence by means of optical (e.g., Soleil-Babinet compensator) or mechanical techniques~\cite{studna:3291}.
However, characterizing the polarization distortion outside of the vacuum with conventional polarimeters is not sufficient to reconstruct separately the birefringence of the two vacuum viewports, which the laser beam must transit.
One solution which has been proposed to obviate this problem requires employing wedged vacuum windows and picking off the beam back-reflected from the inside facet \cite{park2008precision}. However, this is not directly applicable to standard viewports or vacuum cells.

\begin{figure}[t]
	\includegraphics[width=\columnwidth]{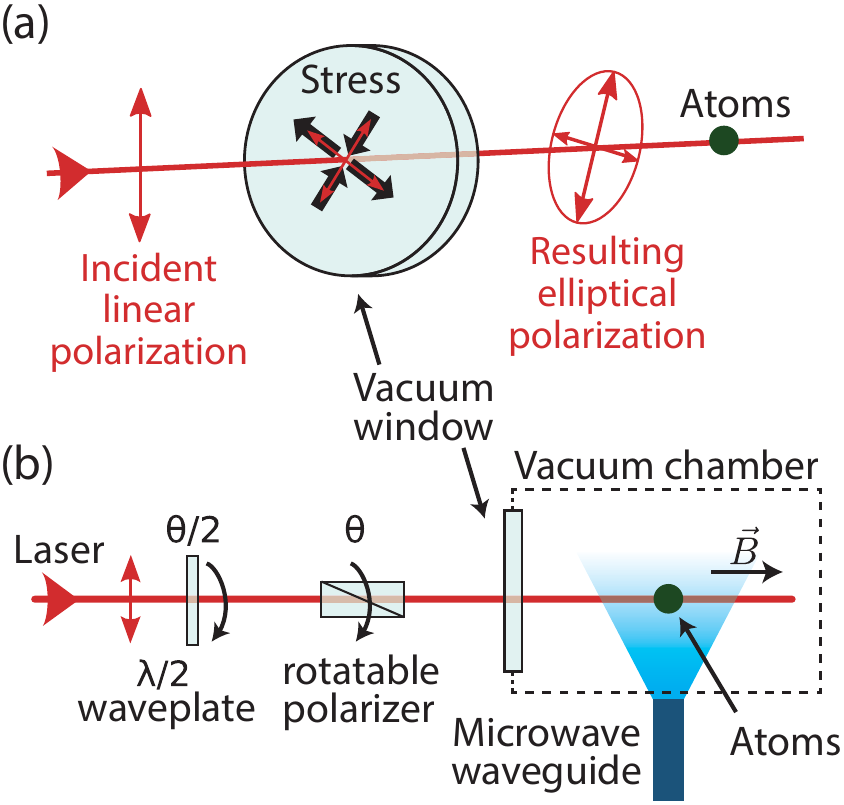}
	\caption{\label{fig:1}a) Polarization distortion caused by stress-induced birefringence in vacuum windows. A linear polarization entering at an angle to the principal stress axes acquires a degree of ellipticity, which in turn induces a detectable shift of a hyperfine transition of the atoms. b) Measurement scheme.
	A half-wave plate and rotatable polarizer allow the linear-polarization of a probe laser beam to be prepared at any arbitrary angle $\theta$.
	While atoms are homogeneously irradiated by the probe laser, the resonance frequency of the hyperfine transition is measured as a function of $\theta$ using microwave spectroscopy.
	The magnetic field $\vec{B}$ defines the quantization axis.}
\end{figure}

In this note, we demonstrate an in-situ method to reconstruct the stress-induced birefringence $\Delta n$ of a vacuum window and the orientation angle $\theta_0$ of the optical axes.
Our scheme makes use of the atoms themselves as a sensitive probe to detect any ellipticity caused by mechanical stresses acting on the vacuum window, as illustrated in figure~\figref{fig:1}{a}.
While varying the angle $\theta$ of the incident linear polarization, we measure the light shift $\delta$ of a hyperfine transition by means of microwave spectroscopy. We will show that the recorded signal behaves as
\begin{equation}
	\label{eq:reference_formula}
	\delta\propto\hspace{1pt}S_0\hspace{1pt}\sin(k\hspace{0.2pt}L\hspace{0.2pt}\Delta n) \sin (2(\theta-\theta_0))\,,
\end{equation}
where the proportionality constant is fully determined by the atomic properties, $k$ is the probe laser wavevector, $L$ the thickness of the vacuum window, and $S_0$ denotes the Stokes parameter, which specifies the total intensity.
Fitting the model (\ref{eq:reference_formula}) to the experimental data allows us to fully reconstruct the Jones matrix
 \begin{equation}
 	M=R(\theta_0)\times
 	\left(\begin{array}{cc}
 	e^{i\hspace{0.2pt}k\hspace{0.2pt}L\hspace{0.2pt}\Delta n/2}&0\\
 	0&e^{-i\hspace{0.2pt}k\hspace{0.2pt}L\hspace{0.2pt}\Delta n/2}
 	\end{array}
 	\right)\times R(-\theta_0)\,,
 	\label{eq:trans_matrix}
 \end{equation}
which characterizes the polarization transformation caused by the vacuum window, where $R(\theta_0)$ is the two-dimensional rotation matrix with angle $\theta_0$. By parametrizing $M$ in the form of (\ref{eq:trans_matrix}), we made the reasonable assumption that mechanical stresses cannot induce optical activity on thin windows \footnote{$M$ represents a SU(2) rotation on the Poincar\'e sphere by an angle $k L \Delta n$ around the axis $(-\sin 2\theta_0,\cos 2\theta_0,0)$, which excludes optical activity.}.  If we represent the incident linear polarization with the Jones vector $\hat{e}=(\cos\theta,\sin\theta)$, the atoms experience the polarization $(u,v)=M\hat{e}$, which has an ellipticity $\epsilon(\theta)=2\operatorname{Im} v^\star u=\sin(2(\theta-\theta_0))\sin(k\hspace{0.2pt}L\hspace{0.2pt}\Delta n)$. We assume here the conventional definition of ellipticity in terms of the Stokes parameters as $\epsilon=S_3/S_0=(I_{\sigma+}-I_{\sigma-})/(I_{\sigma+}+I_{\sigma-})$, with $I_{s}$ being the intensity of $s$-circularly-polarized photons.
Using alkali atoms as an example, we consider the differential vector light shift \cite{deutsch1998quantum,Geremia2006} caused by the probe laser beam to two hyperfine states of the ground state. We call these states $\ket{a}=\ket{F=I-1/2,m_F}$ and $\ket{b}=\ket{F=I+1/2,m'_F}$ in standard spectroscopic notation with the quantization axis in the direction of the probe beam. We obtain from formula (19) in \cite{grimm2000optical} that the resonance frequency is shifted by
\begin{equation}
	\label{eq:light_shift}
	\delta = \alpha \frac{\nu_2-\nu_1}{(\nu-\nu_1)(\nu-\nu_2)}\hspace{0.7pt}(g'_F\hspace{0.5pt}m'_F-g_F\hspace{0.7pt}m_F)\hspace{0.7pt} S_0\hspace{0.7pt}\epsilon\,,
\end{equation}
where $\nu$ is the probe laser frequency, $\nu_1$ and $\nu_2$ denote the resonance frequencies of the D$_1$ and D$_2$ lines, $g_F$ and $g'_F$ represent the $g$-factors of the states $\ket{a}$ and $\ket{b}$, and $\alpha=c^2\Gamma_1/(32\pi^3h\nu_1^3)\approx c^2\Gamma_2/(32\pi^3h\nu_2^3)$ is a proportionality constant depending on the atomic parameters such as the natural decay rates $\Gamma_1$ and $\Gamma_2$ of the doublet states, as well as the speed of light $c$ and Planck constant $h$ \cite{Note2}.\footnotetext{For simplicity's sake, formula (\ref{eq:light_shift}) assumes that the hyperfine interaction is not resolved for sufficiently large detunings. In this approximation, the vector polarizability operator is proportional to the electronic spin \cite{deutsch1998quantum} and, because $L=0$ in the ground state and because of the rotational symmetry along the quantization axis, it can be expressed through $g_F\hspace{0.5pt}m_F$ using the Land\'e projection theorem. Here, $g_F$ does not include the nuclear contribution.}
Inserting the expression of the ellipticity $\epsilon(\theta)$ in (\ref{eq:light_shift}), we readily obtain the expression in (\ref{eq:reference_formula}).
Formula (\ref{eq:light_shift}) prompts a few considerations about the sensitivity $\beta=\delta/(\epsilon\,S_0)$ of atoms to elliptical polarization per unit of laser intensity:
The sensitivity is proportional to the line doublet splitting $\nu_2-\nu_1$ (in general, spin-orbit coupling), which makes heavier atoms more sensitive than lighter atoms. In addition, probe frequencies $\nu$ close to the optical resonance are more favorable, since the sensitivity scales approximately with $1/(\nu-\nu_0)^2$ for large detunings. But more importantly, a pair of magnetic-field sensitive states $\ket{a}$ and $\ket{b}$ must be chosen to measure the vector light shift, so that $g'_Fm'_F\neq g_Fm_F$; this excludes, for instance, employing the so-called clock states.

\begin{figure}[t]
	\includegraphics[width=\columnwidth]{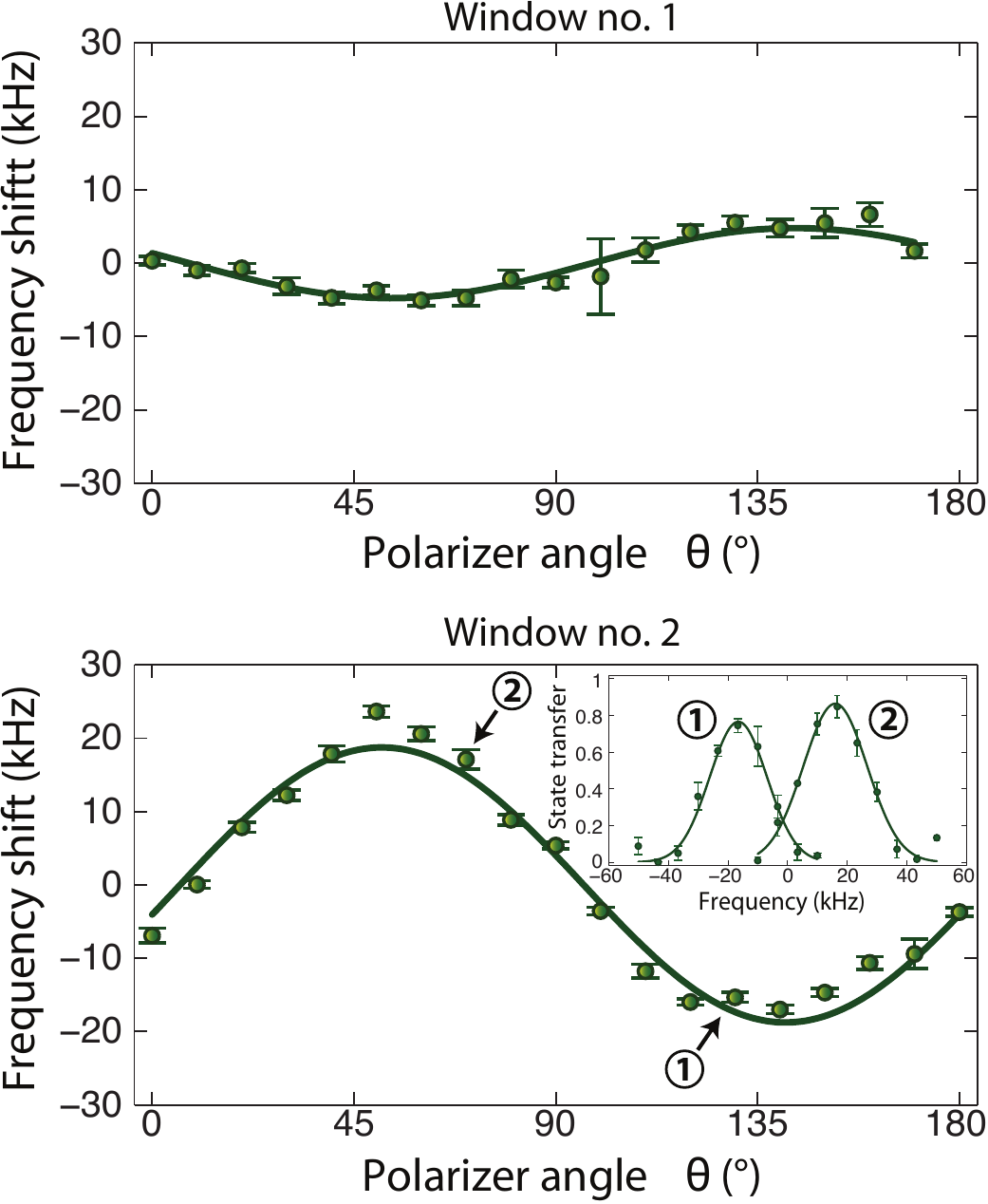}
	\caption{\label{fig:2}Measured differential light shift as a function of the polarization angle $\theta$. The transition frequency varies in a sinusoidal way according to (\ref{eq:reference_formula}). Shown are the data with 1 sigma error bars and a sinusoidal fit. The offset of $\SI{9.199810}{GHz}$, including the contribution from differential scalar light shift and magnetic fields, has been subtracted. The origin $\theta=0$ is arbitrarily set by the polarizer's orientation in its mount. Inset: Two measured spectra, corresponding to the two indicated data points.}
\end{figure}

\begin{table*}[t]
\centering
\caption{Measurement results with statistical errors. From top to bottom, the maximum transition shift (for $\theta=\theta_0+\ang{45}$) and the zero-crossing angle are reported. Inferred from $\delta_0$ are the maximal ellipticity and the magnitude of birefringence, with the systematic error coming from the uncertainty on the total light intensity.}
\begin{tabular}{c@{\hspace{0.5cm}}c@{\hspace{0.5cm}}c}
	\\[-2.5mm]
	&Window no.~1&Window no.~2\\[0.1mm]
	\begin{tabular}{@{\hspace{2mm}}l@{\hspace{-3mm}}l@{\hspace{-0.5mm}}l@{\hspace{1mm}}}
		\hline\\[-3.2mm]
		Quantity\\[0.3mm]
		\hline\\[-3mm]
		$\delta_0$\hspace{1mm}(kHz)\\
		$\theta_0$\hspace{1mm}($^\circ$)\\[0.5mm]
		\hline\\[-3mm]
		$\operatorname{max} \epsilon$ &&$(10^{-3})$ \\
		$\Delta n$&&$(10^{-7})$\\[0.5mm]
		\hline
	\end{tabular}
	&\begin{tabular}{l@{\hspace{0.5cm}}l@{\hspace{0.5cm}}l}
	\hline\\[-3.2mm]
	Value&Statistical&Systematic \\[0.3mm]
	\hline\\[-3mm]
	$4.8$ & $\pm0.4$ & -- \\
	$98$ &$\pm2$& -- \\[0.5mm]
	\hline\\[-3mm]
	$8.2$ & $\pm0.7$ & $\pm0.9$ \\
	$2.3$ & $\pm0.2$ & $\pm0.2$  \\[0.5mm]
	\hline
\end{tabular}&
\begin{tabular}{l@{\hspace{0.5cm}}l@{\hspace{0.5cm}}l}
	\hline\\[-3.2mm]
	Value&Statistical&Systematic \\[0.3mm]
	\hline\\[-3mm]
	$18.8$ & $\pm0.9$ & -- \\
	$6$ &$\pm1$& -- \\[0.5mm]
	\hline\\[-3mm]
	$64$ & $\pm3$ & $\pm6$ \\
	$17.7$ & $\pm0.8$ & $\pm1.8$  \\[0.5mm]
	\hline
\end{tabular}
\end{tabular}
\label{tab:res}
\end{table*}

In our experiment, a small ensemble of ${}^{133}$Cs atoms is cooled in a magneto-optical trap and subsequently transferred to an optical dipole trap. The trap stems from the probe laser beam itself, which has a wavelength of $\SI{866}{\nano\meter}$ and is tightly focused on the atoms.
The light polarization is precisely set by a rotatable Glan-laser polarizer ($\ang{0.5}$ angle reproducibility) with a preceding half-wave plate used to maximize the transmission through the polarizer, as shown in figure~\figref{fig:1}{b}; the purity of the incident polarization is determined by the polarizer's extinction ratio of $\num{e-6}$.
We tested two different vacuum windows: a cuboid glass cell made of Corning Vycor$^\text{\textregistered}$ 7913 (manufactured by Hellma) and a standard CF63 viewport of made of Spectrosil$^\text{\textregistered}$ 2000.
We will call them, for convenience, window no.~1 and no.~2.
Both are \SI{5}{mm} thick, while the probe intensity experienced by the atoms during the microwave spectroscopy is $\sim\SI{12}{W/mm^2}$ for window no.~1 and $\sim\SI{6}{W/mm^2}$ for window no.~2.
A magnetic field of about \SI{3}{Gauss} is added along the probe beam's direction to define the quantization axis, and the atoms are prepared by optical pumping to the state $\ket{b}=\ket{F=4,m_F=4}$.
We drive the transition to the state $\ket{a}=\ket{F=3,m_F=3}$ using microwave pulses at varying frequencies around $\SI{9.2}{GHz}$.
With this choice of the Zeeman states, we obtain the largest sensitivity to elliptical polarization (as $g'_F=-g_F$), resulting in $\beta=-\SI{50}{\kilo\hertz/(\watt/\mm^2)}$ for our case. The recorded Fourier-limited spectra exhibits a FWHM of about $\SI{30}{kHz}$ \cite{meschede2006manipulating}.

We obtain the light shift $\delta$ from the recorded spectra as the displacement of the center frequency for different angles $\theta$, see figure~\ref{fig:2}.
The data exhibit the expected sinusoidal behavior $\delta=\delta_0\sin(2(\theta-\theta_0))$ with the amplitude specified by $\delta_0=\beta S_0\sin(kL\Delta n)$.
The zero-crossing determines the angle $\theta_0$ of one of the optical axes, while the inferred value of $\delta_0$ yields the birefringence $\Delta n$.
The results are listed in table~\ref{tab:res}, showing that the birefringence magnitude and the orientation of optical axes can be extracted with good precision. We attain in our case a precision level of $\num{e-8}$ even though our apparatus can only probe small ensembles of less than $\num{30}$ atoms at each iteration (\SI{2}{\s} cycle time).
Because the light shift $\delta_0$ is proportional to $S_0$, we must take into account a conservative $10\%$ systematic uncertainty arising from the probe beam's intensity, which depends on the precise position and geometry of the beam waist.
If required, $S_0$ can be precisely calibrated by introducing in the incident polarization a known amount of ellipticity (up to $\epsilon=1$) and measuring the resulting light shift.
From our measurements, we learn that the glass cell (window no.~1) exhibits a smaller birefringence than the commercial viewport (window no.~2).
A possible explanation is that the frontal surface of the cell is further away from the mounting flange, where most of mechanical stresses are localized \cite{solmeyerVacuum}.
If we treat the window and its internal stresses as planar and assume the stress distribution to be homogeneous over the size of the laser beam, the stress distribution can be expressed in terms of two orthogonal principal stresses $\Sigma_1, \Sigma_2$, as illustrated in figure~\figref{fig:1}{a}. We can hence use the stress optic law to derive the differential stress $\Sigma_\text{D}= \Sigma_1-\Sigma_2=\Delta n/C$ from the measured birefringence $\Delta n$, with $C$ being the stress optic coefficient of the material \cite{born1999principles}.
In case of fused silica, which is the main constituent of our windows, $C$ is around $\SI{3.5e-12}{\per\pascal}$, yielding a $\Sigma_\text{D}$ of $\SI{70}{\kilo\pascal}$ for window no.~1 and $\SI{500}{\kilo\pascal}$ for window no.~2.
These are typical values occurring in a vacuum cell and a commercial Conflat viewport, respectively \cite{studna:3291}. In critical applications demanding ultralow stress-induced birefringence, a special glass like Schott SF57 could provide two orders of magnitude suppression due to its unusually small stress-optic coefficient of $\SI{2e-14}{\per\pascal}$ \cite{McMillan2004}.

In conclusion, we have presented a novel in-situ method to measure with high precision the ellipticity caused by the stress-induced birefringence.
This method allows us to align the incident linear polarization along one of the optical axes,  virtually canceling the effect of birefringence.
Alternatively, one could employ a waveplate with the retardance set to $kL\Delta n$ or a Soleil-Babinet compensator in order to preserve any arbitrary incident polarization.

\begin{acknowledgments}
	We acknowledge financial support from DFG Research Unit FOR 635, NRW-Nachwuchsforschergruppe ``Quantenkontrolle auf der Nanoskala'', ERC grant DQSIM, EU project SIQS, BCGS program, and Studienstiftung des deutschen Volkes. AA also acknowledges support from the Alexander von Humboldt Foundation.  
\end{acknowledgments}

\ifusebibfile

\bibliography{bibref}

\else


%


\fi

\end{document}